\documentclass[a4paper]{article}
\usepackage{amsmath,amssymb,bm,epsfig}
\begin{document}

\newcommand{\ba}{\mbox{\boldmath $a$}}
\newcommand{\bb}{\mbox{\boldmath $b$}}
\newcommand{\bc}{\mbox{\boldmath $c$}}

\title{Approximate vortex solution of Faddeev model}

\author{Chang-Guang Shi\\Department of  Mathematics and Physics, \\Shanghai
University of Electric power\\
 Shanghai,200090 {\bf China}\\
shichangguang@shiep.edu.cn\\
\\
{Minoru Hirayama} \\Department of Mathematics and Physics,\\
Shanghai
University of Electric power\\
Shanghai,200090 {\bf China}\\
Department of Physics,  University of Toyama, Gofuku 3190\\
 Toyama, 930-8555, {\bf Japan}\\
hirayama@jodo.sci.u-toyama.ac.jp}
\date{}
\maketitle


\begin{abstract}
Through an Ansatz specifying the azimuthal-angle dependence of the solution,
 the static field equation for vortex of the Faddeev model is converted to an algebraic ordinary differential equation.
 An approximate analytic expression of the vortex solution is explored so that the energy per unit vortex length becomes as small as possible.
 It is observed that the minimum energy of vortex is approximately proportional to the integer which specifies the solution.
\end{abstract}


\section{Introduction}

Faddeev model\cite{Faddeev} was originally proposed as a model which
might give rise to 3-dimensional soliton solutions. Later it was
discussed that the model might be an effective field theory
describing the low energy behavior of the $SU(2)$ gauge
field\cite{FN}. It can be regarded as a restricted version of the
Skyrme model\cite{Sky} which may be an effective field theory for
hadron dynamics. Because of the high nonlinearity of Faddeev and
Skyrme models, the analytic structures of solutions of these models
have not yet been clarified. The numerical solutions of these
models, however, exhibit quite interesting soliton
properties\cite{Ba-Sut1}-\cite{Hi-So}: soliton solutions of the
Skyrme model have tetrahedral structures while those of the
 Faddeev model have knot structures\cite{Ba-Sut1},\cite{Ba-Sut2}.

The Faddeev model is a model concerning the real scalar fields
\begin{equation}
\boldsymbol{n}(x)=\left(n^1(x),n^2(x),n^3(x)\right)
\end{equation}
 satisfying
\begin{equation}
{\boldsymbol{n}}^2(x)=\boldsymbol{n}(x)\cdot\boldsymbol{n}(x)=\sum\limits_{a=1}^{3}n^a(x) n^a(x)=1.
\end{equation}
It is defined by the Lagrangian density
\begin{align}
{\mathcal L}_F(x)&=c_2 l_2(x)+c_4 l_4(x),\\
l_2(x)&=\partial_{\mu}\boldsymbol{n}(x)\cdot
 \partial^{\mu}\boldsymbol{n}(x),\\
l_4(x)&=-H_{\mu\nu}(x) H^{\mu\nu}(x),\\
H_{\mu\nu}(x)&=\boldsymbol{n}(x)\cdot[\partial_{\mu}\boldsymbol{n}(x)\times
 \partial_{\nu}\boldsymbol{n}(x)]\nonumber\\
&=\epsilon_{abc}n^a(x)\partial_{\mu}n^b(x)\partial_{\nu}n^c(x),
\end{align}
where $c_2$ and $c_4$ are constants. The static energy
functional $E_F[\boldsymbol{n}]$ associated with ${\mathcal L}_F(x)$ is
given by
\begin{align}
E_F[\boldsymbol{n}]&=\int dV \epsilon(\boldsymbol{x}),\\
 \epsilon(\boldsymbol{x})&=c_2 \epsilon_2(\boldsymbol{x})+c_4
 \epsilon_4(\boldsymbol{x}),\\
\epsilon_2(\boldsymbol{x})&=\sum\limits_{a=1}^{3}\sum\limits_{i=1}^{3}[\partial_i
 n^a(\boldsymbol{x})]^2,\\
\epsilon_4(\boldsymbol{x})&=\sum\limits_{i,j=1}^{3}[H_{ij}(\boldsymbol{x})]^2,
\end{align}
with $\boldsymbol{x}=(x_1,x_2,x_3)$ and $dV=dx_1 dx_2 dx_3$.
By the stereo-graphic projection, the field $\bm{n}$ can be expressed by a complex function $u$ as
 \begin{equation}
{\boldsymbol n}=\biggl(\frac{u+u^{*}}{|u|^2+1},\frac{-i(u-u^{*})}{{|u|}^2+1},\frac{{|u|}^2-1}{{|u|}^2+1}\biggr).
\end{equation}
In terms of $u$, the energy densities $\epsilon_2$ and $\epsilon_4$ are given by
\begin{equation}
\epsilon_2=\frac{4}{(1+|u|^2)^2}(\boldsymbol{\nabla} u\cdot\boldsymbol{\nabla} u^{*}),
\end{equation}
\begin{align}
\epsilon_4=-8\frac{({\boldsymbol \nabla}u\times{\boldsymbol \nabla}u^*)^2}{(1+|u|^2)^4}.
\end{align}
The field equation can be rather simply expressed by $\bm{q}$ defined by
\begin{align}
\bm{q}=X\nabla u,
\end{align}
with
\begin{align}
u=R{\rm{e}}^{i\Phi},\quad R=|u|
\end{align}
and
\begin{align}
X=2\sqrt{\frac{c_4}{c_2}}\frac{1}{1+R^2}=\frac{1}{1+R^2},
\end{align}
where $2\sqrt{c_4/c_2}$ of the dimension of length has been set equal to $1$ and $\bm{q}$ is dimensionless. The static field equation can be
written as\cite{Hi-Shi}
\begin{align}
\nabla\cdot\bm{\alpha}+i \bm{\beta}\cdot\bm{\alpha}=0,
\label{eqn:alphaEq}
\end{align}
where $\bm{\alpha}$ and $\bm{\beta}$ are a complex and a real $3$-vectors given by
\begin{align}
\bm{\alpha}&=\bm{q}^{\star}-\bm{q}^{\star}\times(\bm{q}\times\bm{q}^{\star}),\label{eqn:alphaDEF}\\
\bm{\beta}&=\frac{1}{i}(u^{\star}\bm{q}-u\bm{q}^{\star})= B\nabla
\Phi,~B=\frac{2R^2}{1+R^2},
\end{align}
respectively.

In this paper, we discuss some simple analytic solutions of the
above equation. We shall consider mainly the vortex solutions of the
form $R=R(\rho),~\Phi=m\phi$ with $(\rho, z, \phi)$ and $m$ being
the cylindrical coordinates and an integer, respectively.
 We adopt an approximate analytic solution containing two adjustable parameters. Its form is  fixed by the compatibility with the boundary condition and the singularity-structure necessitated by the field equation. The parameters are fixed so that the energy per unit length of vortex, $A_m$, becomes minimal. We find that the minimum of $A_m$ is proportional to $m$ approximately. \\
\hspace{3mm} This paper is organized as follows. In Sec.2, we discuss the $m=0$ case briefly. In Sec.3, the above-mentioned approximate analytic solution for the $m\neq 0$ case is discussed. Sec.4 is devoted to summary.

\section{Some solutions in special cases}

\hspace{3mm}We first consider the solutions of the form
\begin{align}
R=R(\rho, z),\quad \Phi=\Phi(\phi),
\end{align}
where $(\rho, z, \phi)$ are cylindrical coordinates.
Then we have
\begin{align}
\bm{q}&=X {\rm{e}}^{i\Phi} \left(R_{\rho}\bm{e}_{\rho}+R_{z}\bm{e}_{z}+\frac{iR\Phi^{'}}{\rho }\bm{e}_{\phi} \right),\\
\bm{\alpha}&= {\rm{e}}^{-i\Phi}\left(C\bm{e}_{\rho}+D\bm{e}_{z}+F\bm{e}_{\phi} \right),\\
\bm{\beta}&=\frac{B \Phi^{'}}{\rho}\bm{e}_{\phi},\\
C&=\left(X+\frac{R^2Y (\Phi^{'})^2}{\rho^2}\right)R_{\rho},\\
D&=\left(X+\frac{R^2Y (\Phi^{'})^2}{\rho^2}\right)R_{z},\\
F&=-i\frac{R \Phi^{'}}{\rho}\left[X+Y\left(R_{\rho}^2+R_z^2 \right)\right],\\
Y&=2X^3,\\
\Phi^{'}&=\frac{d\Phi(\phi)}{d\phi},\quad  R_{\rho}=\frac{\partial R(\rho,z)}{\partial\rho}, \quad R_{z}=\frac{\partial R(\rho,z)}{\partial z},
\end{align}
where $\bm{e}_{\rho}, \bm{e}_{z},$ and  $\bm{e}_{\phi}$ are  orthonormal  unit vectors satisfying $\bm{e}_{\rho} \times \bm{e}_{\phi}=\bm{e}_{z}$, etc.\\
We obtain $F_{\phi}=0$ from Im\{$\nabla\cdot\bm{\alpha}+i
\bm{\beta}\cdot\bm{\alpha}$\}=$0$. Then, taking the
single-valuedness of $u$ into account, we are led to
\begin{equation}
\Phi=m\phi,\quad m:\rm{integer}.
\end{equation}
On the other hand, from Re\{$\nabla\cdot\bm{\alpha}+i \bm{\beta}\cdot\bm{\alpha}$\}=$0$ and $\Phi^{'}$=$m$, we have
\begin{align}
\nabla\cdot &\left(XG\nabla R\right)+(G-1)\frac{R^2-1}{2R}\left[1+2X^2\left(\nabla R\right)^2\right]=0,\\
G&=1+\frac{2m^2R^2X^2}{\rho^2}.\label{eqn:2.12}
\end{align}

In the $m=0$ case, it becomes $\triangle[{\rm{arctan}}R]=0$, whose solution is given by
\begin{align}
{\rm{arctan}}R&=\sum_{n=0}^{\infty}\left(A_n r^n+\frac{B_n}{r^{n+1}}\right){\rm{P}}_n(\rm{cos} \theta),\\
r&=\sqrt{\rho^2+z^2}, \quad {\rm{cos}}\theta=\frac{z}{r},
\end{align}
where $\{A_n, B_n :n= 0, 1, 2, \cdots\}$ are constants and ${\rm{P}}_n$ denotes the $n$-th Legendre polynomial.
It should be noted that
\begin{equation}
{\rm{arctan}}R= a_1~{\rm{ln}}\rho+a_2 \quad (a_1, a_2: \rm{const.})
\end{equation}
is a solution, too.
These results indicate how Cho's solution\cite{Cho} $u=\frac{\rm{const.}}{r}$ should be generalized. They are, however, trivial in the sense that the interference effects between  the $l_2(x)$ and  $l_4(x)$ in the Lagrangian disappear in these solutions.\\
\hspace{3mm} For $m\neq 0$, we see that the simpler cases such as
$R=f_1(z), R=f_2(r),~R=f_3(\theta),~R=f_4(\eta),~R=f_5(\xi)$ are not
allowed since $G$ in Eq. (\ref{eqn:2.12}) contains $\rho$, where
$(r,\theta, \phi)$ and $(\eta, \xi, \phi)$ are polar- and toroidal-
coordinates, respectively. Only the vortex solution $R=R(\rho)$,
which is a solution of the $2$-dimensional Faddeev model, is
allowed.

\section{Approximate vortex solutions}
\hspace{3mm}We hereafter consider the case $R=R(\rho)$ and $m\neq 0$.
If we set
 \begin{align}
 R={\rm{tan}}\frac{\xi}{2},\quad \sigma=\frac{\rho}{m},
 \end{align}
we obtain
\begin{align}
 & \left(\sigma^2+2\rm{sin}^2\xi \right)\frac{d^2 \xi}{d\sigma^2}+{\rm{sin}}2\xi\left(\frac{d\xi}{d\sigma}\right)^2+\left(\sigma-\frac{2}{\sigma}{\rm{sin}}^2\xi\right)\frac{d\xi}{d\sigma}-2m^2{\rm{sin}}2\xi=0.
 \end{align}
Through a further change of variables
 \begin{align}
&V(\zeta)=-{\rm{tan}}^2\xi=-\left(\frac{2R}{R^2-1}\right)^2,\quad
\zeta=\frac{\sigma^2}{\sigma^2+2}
 \end{align}
we obtain an algebraic differential equation
 \begin{align}
&\frac{d^2V}{d\zeta^2}-\frac{1}{2}\left(\frac{1}{V}+\frac{1}{\zeta-V}+\frac{3}{V-1}\right)\left(\frac{dV}{d\zeta}\right)^2 \nonumber \\
&\hspace{2cm} +\left(\frac{1}{\zeta-1}+\frac{1}{\zeta-V}\right)\frac{dV}{d\zeta}-\frac{2m^2 V(V-1)}{\zeta(1-\zeta)^2(V-\zeta)}=0.
 \end{align}
 The solution of this equation can be explored in the following way. For a given $\zeta_0$ which is different from $0$ and $1$,  we assume that the solution $V(\zeta)$ near $\zeta=\zeta_0$ is given as
 \begin{align}
&V(\zeta)=\sum_{n=0}^{\infty}V_n \left(\zeta-\zeta_0\right)^{\alpha+n},\quad 0<\zeta_0<1,\quad \alpha<0.
\end{align}
Then we are led to
\begin{align}
&\alpha=-2
\end{align}
and
\begin{align}
&V_1=\frac{V_0}{\zeta_0-1},\\
&V_2=\frac{(4m^2+\zeta_0) V_0-4\zeta_0(\zeta_0-1)^2(\zeta_0-3)}{12\zeta_0(\zeta_0-1)^2},\\
&V_3=-\frac{m^2 V_0+2
\zeta_0^2(\zeta_0-1)^2}{6\zeta_0^2(\zeta_0-1)^2},
\end{align}
and so on.
It turns out that $V_0$ and $\zeta_0$ are arbitrary and $V_1, V_2, V_3, \cdots$ are fixed by them.\\
Although $V_n$ can be determined order by order, it is difficult to conclude
that the series (3.5) converges in some domain of $\zeta$ around $\zeta_0$.
To obtain the radius of convergence of the series (3.5), if any, we must calculate
$V_n$ for very large $n$, whose mathematical expression becomes quite
complicated as $n$ gets large.
The differential equation (3.4) suggests that the behavior of $V(\zeta)$ at
 $\zeta=0$ and $1$ ($\rho=0$ and $\infty$) may be different from that at ordinary
  points $0<\zeta<1$ ($0<\rho<\infty$).
It is also difficult to maintain the properties of $V(\zeta)$ near $\zeta
=0$ and $1$  by an infinite series of the form (3.5).
We therefore consider an approximate $V(\zeta)$ which realizes the properties
required by the lowest order analysis of (3.4) around $\zeta=\zeta_0~(\neq0,1)$,
$\zeta=0$ and $\zeta=1$.\\
As was explained above, the behavior of $V(\zeta)$ in the neighborhood of $\zeta_0$
is given by $V(\zeta)\sim {\rm{const.}}(\zeta-\zeta_0)^{-2}$.
We next consider the behavior of $V(\zeta)$ near $\zeta=0$.
It turns out that the behavior $V(\zeta)\sim$const.$\zeta^\lambda$ is compatible
with (3.4) only when $\lambda=2m$ or $\lambda=-2m$.
We here recall that, with the assumption (2.10), only $R=0$ and $R=\infty$ yield
the vanishing energy density irrespective of the value of $\phi$.
These values of $R$ imply $V(0)=0$, leading to $\lambda=2|m|$.
In other words, if we require that the energy density is vanishing at $\rho=0$,
the differential equation (3.4) and the assumption (2.10) lead us to
\begin{align}
V(\zeta)\sim \rm{const.} \zeta^{2|m|}, \quad \zeta\sim 0.
 \end{align}
 Similarly, from the requirement that the energy density is vanishing at $\rho=\infty$ , we obtain
\begin{align}
V(\zeta)\sim \rm{const.} (\zeta-1)^{2|m|}, \quad  \zeta\sim 1.
 \end{align}
It is straightforward to see that the conditions $V(0)=V(1)=0$ correspond to
the configurations $\boldsymbol{n}=(0, 0, 1)$ or $\boldsymbol{n}=(0, 0, -1)$
at $\rho=0$ and $\infty$. Therefore we are here considering the configurations
 interpolating these configurations. In the example considered below (Fig.4),
 we obtain $V(\zeta)$ connecting   $\boldsymbol{n}=(0, 0, 1)$ at $\rho=0$ and
 $\boldsymbol{n}=(0, 0, -1)$ at $\rho=\infty$.

We note that, in the analysis of the hedgehog Skyrmion, the solution of the differential equation
 \begin{align}
&\frac{d^2W}{d\eta^2}-\frac{1}{2}\left(\frac{1}{W}+\frac{1}{\eta-W}+\frac{3}{W-1}\right)\left(\frac{dW}{d\eta}\right)^2 \nonumber \\
&\hspace{2cm} +\left[\frac{1}{2}\left(\frac{1}{\eta-1}+\frac{1}{\eta}\right)+\frac{1}{\eta-W}\right]\frac{dW}{d\eta}-\frac{W[(\eta+1)W-2\eta]}{2\eta^2(\eta-1)^2(W-\eta)}=0
 \end{align}
was investigated\cite{Ya-Hi} by a trial function
\begin{equation}
W=\frac{-q\eta(\eta-1)^2}{\eta-p}, \quad q>0, \quad 1>p>0,
\end{equation}
which yielded a rather good value of the energy. \\

In the following, we consider the approximate solutions $V_m(\zeta)$ given by
 \begin{equation}
 V_m(\zeta)=-q_m\frac{\left[\zeta(1-\zeta)\right]^{2m}}{(\zeta-p_m)^2}, \quad m > 0, \quad q_m>0,~\quad 1>p_m>0,
 \end{equation}
 which is compatible with the singularity-structure indicated by the differential equation and the boundary condition that the energy density vanishes at $\rho=0, \infty$. \\
The energy $E_F$ corresponding to the assumption (2.10), which we denote by $E_m$, is now given by
 \begin{align}
 E_m &=8\pi c_2\int_{-\infty}^{\infty} A_m dz,\\
 A_m &=\int_0^1 W_m d\zeta,\\
 W_m &=\frac{m^2}{8}\frac{q_m\left[\zeta(1-\zeta)\right]^{2m-1}}{(\zeta-p_m)^2+q_m\left[\zeta(1-\zeta)\right]^{2m}} \nonumber\\
 &+\frac{1}{2}\frac{q_m\left[\zeta(1-\zeta)\right]^{2m-1} \left[\zeta(1-\zeta)+m(2\zeta-1)(\zeta-p_m)\right]^2} {\left\{(\zeta-p_m)^2+q_m\left[\zeta(1-\zeta)\right]^{2m}\right \}^2} \nonumber\\
 &+\frac{1}{8}\frac{q_m^2 \zeta^{4m-2}(1-\zeta)^{4m} \left[\zeta(1-\zeta)+m(2\zeta-1)(\zeta-p_m)\right]^2} {\left\{(\zeta-p_m)^2+q_m\left[\zeta(1-\zeta)\right]^{2m}\right \}^3}.
 \end{align}
 We fix $p_m$ and $q_m$ so that they minimize $A_m$. For $m=1\sim 4$, they are given as\\

$ \begin{cases}
 &m=1:\quad  p_1=0.83,\quad q_1=4.0, \quad A_1=1.14\\
 &m=2:\quad  p_2=0.61,\quad q_2=2.3\times 10, \quad A_2=2.23\\
 &m=3:\quad  p_3=0.56,\quad q_3=1.57\times  10^2, \quad A_3=3.28\\
 &m=4:\quad  p_4=0.52,\quad q_4=1.36\times 10^3, \quad A_4=4.34.\\
 \end{cases}$\\

 It seems that $p_m$ approaches to $0.5$ when $m$ becomes large. For the convenience of numerical estimation, we fix $p_m$ as $0.5$ for $m$ larger than $4$ and determine $q_m$ so as to minimize $A_m$. The result is given as follows:\\

 $
 \begin{cases}
 &m=5: \quad q_5=1.35\times 10^4 \quad A_5=5.42\\
 &m=6: \quad q_6=1.45\times 10^5 \quad A_6=6.49\\
 &m=7: \quad q_7=1.66\times 10^6 \quad A_7=7.57\\
 &m=8: \quad q_8=1.96 \times 10^7 \quad A_8=8.65\\
 &m=9: \quad q_9=2.41\times 10^8 \quad A_9=9.74\\
 &m=10:\quad q_{10}=3.06 \times 10^9 \quad A_{10}=10.8.\\
  \end{cases}
$ \\
The results for $A_m$ and $q_{m+1}/q_m$  are given in Fig.1 and Fig.2, respectively.
We are then led to the approximate formulae
\begin{equation}
A_m=a+bm,\quad a=0.052,~b=1.076
\end{equation}
and
\begin{equation}
\frac{q_{m+1}}{q_m}=c\left(1-d~{\rm{e}}^{-fm}\right), \quad c=14.0,~ d=0.769,~f=0.234.
\end{equation}

\begin{figure}[pb]
\centering \psfig{file=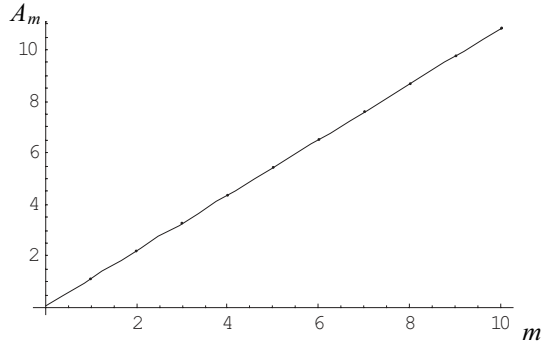,width=7cm}\caption{$A_m$ vs. $m$}
\label{fig:1}
\end{figure}

\begin{figure}[tpb]
\centering\psfig{file=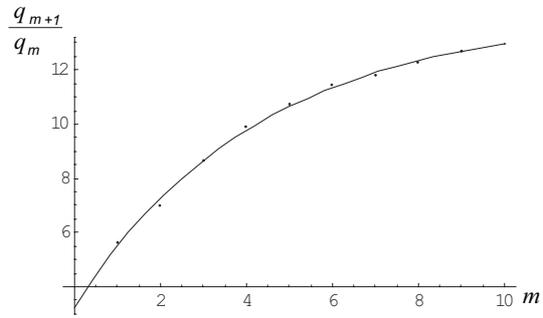,width=7cm}
\caption{$\frac{q_{m+1}}{q_m}$ vs. $m$} \label{fig:2}
\end{figure}


Since the infinite product $\prod_{m=0}^{\infty}\left(1-d~{\rm{e}}^{-fm}\right)\equiv B$ converges to $0.00592$,
 we have ${\rm{log}}~q_m\sim m~{\rm{log}}~c +{\rm{log}}~B$ for large $m$.
 From the fact that $a\sim 0$  and $b \sim A_1$, it looks like that there exists an atom-like object Q with  energy $b$
  and unit $m$ and that the assembly of $m$ Q's constitutes the  configuration with $m$.

 With the original variables, the energy density $K(\rho)$ defined by $A_1=\int_0^{\infty} K(\rho) d\rho$ and $R(\rho)$ for
 $m=1$ are given in Fig.3 and Fig.4, respectively.

\begin{figure}[pb]
\centerline{\psfig{file=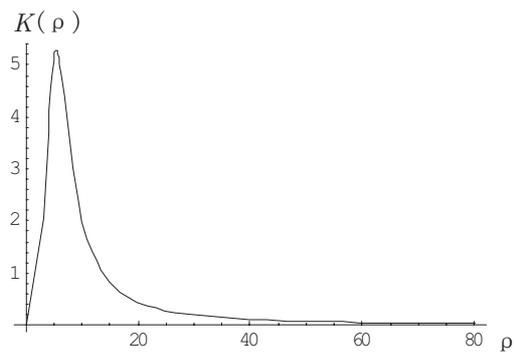,width=7cm}}\caption{Energy density
$K(\rho)$ vs. $\rho$} \label{fig:3}
\end{figure}

\begin{figure}[pb]
\centering\psfig{file=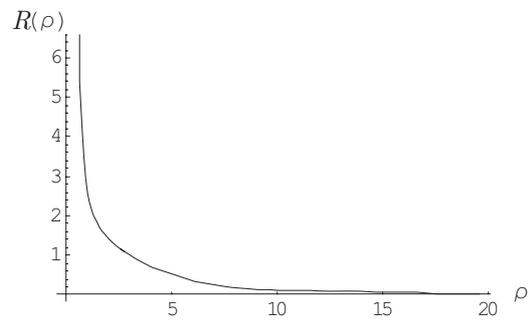,width=7cm} \caption{$R(\rho)$ vs.
$\rho$} \label{fig:4}
\end{figure}


\section{Summary}
We have examined the vortex solutions of the Faddeev model. Through
the change of variables, the field equation was converted to an
algebraic differential equation containing an integer-parameter $m$.
Its approximate solution was parameterized with the aid of two
parameters $p_m$ and $q_m$. They were fixed so that the energy per
unit length of vortex, $A_m$, became minimal. It was observed that
the minimum of $A_m$ is proportional to $m$ approximately.
 The numerical analysis of the Faddeev model made so far clarified
 the knot structure of the genuine three-dimensional solitons.
  We hope that our two-dimensional analysis might be helpful for
   the understanding of the latter since knot-solitons may be regarded
    as the bended and twisted vortices.

\section*{Acknowledgments}
This research was partially supported by the National Natural
Science Foundation of China (Grant No. 10601031).


\begin{thebibliography}{0}
\bibitem{Faddeev}L. Faddeev,
{\it Lett. Math. Phys. }{\bf 1}, (1976) 289.
\bibitem{FN}L. Faddeev and A. J. Niemi,
{\it Phys. Rev. Lett. }{\bf 82}, (1999)  1624.
\bibitem{Sky}T.H.R. Skyrme,
{\it Nucl. Phys.}{31} (1961) 556.
\bibitem{Ba-Sut1}R. A. Battye and P. M. Sutcliffe,
{\it Phys. Rev. Lett. }{\bf 81}, (1998) 4798.
\bibitem{Ba-Sut2}R. A. Battye and P. M. Sutcliffe,
{\it Phys. Rev. Lett. }{\bf 79}, (1997) 363.
\bibitem{Hi-So}J. Hietarinta and P. Salo,
{\it Phys. Lett. B} {\bf  451},  (1999) 60.
\bibitem{Hi-Shi}M. Hirayama and C.-G. Shi,
{\it Phys. Lett. B}{\bf 652},(2007) 384.
\bibitem{Cho}Y. M. Cho,
Phys. Rev. Lett. {\bf{87}}, 252001 (2001).
\bibitem{Ya-Hi}J. Yamashita and M. Hirayama,
{\it Phys. Lett. B}{\bf 642},(2006) 160.
\end{thebibliography}
\end{document}